\documentclass[aps, pra,showpacs,superscriptaddress,floatfix,english,twocolumn]{revtex4-2}

\usepackage{graphicx}
\usepackage{dcolumn}
\usepackage{bm}
\usepackage{hyperref}

\usepackage{xcolor}
\usepackage{mathrsfs}
\usepackage{subfigure}

\usepackage{physics}
\usepackage[page]{appendix}

\hypersetup{
    unicode=false,          
    pdftoolbar=false,        
    pdfmenubar=true,        
    pdffitwindow=false,     
    pdfstartview={FitH},    
    pdfkeywords={MBL} {AGP} {Mass-deformed SYK}, 
    pdfnewwindow=true,      
    colorlinks=true,       
    linkcolor=red,          
    citecolor=blue,        
    filecolor=magenta,      
    urlcolor=blue           
}

\newcommand{\hH}{\hat{\mathcal{H}}}

\newcommand{\rr}{\langle r \rangle}

\newcommand{\ictpadd}{ICTP South American Institute for Fundamental Research \\
Instituto de F\'{i}sica Te\'{o}rica, UNESP - Univ. Estadual Paulista \\
Rua Dr. Bento Teobaldo Ferraz 271, 01140-070, S\~{a}o Paulo, SP, Brazil}
\newcommand{\pcsadd}{Center for Theoretical Physics of Complex Systems, Institute for Basic Science (IBS), Daejeon - 34126, Korea}
\newcommand{\ustadd}{Basic Science Program, Korea University of Science and Technology (UST), Daejeon - 34113, Korea}

\begin{document}

\title{From Dyson Models to Many-Body Quantum Chaos}

\author{Alexei Andreanov}
    \email{aalexei@ibs.re.kr}
    \affiliation{\pcsadd}
    \affiliation{\ustadd}

\author{Matteo Carrega}
    \email{matteo.carrega@spin.cnr.it}
    \affiliation{CNR-SPIN, Via Dodecaneso 33, 16146 Genova, Italy}

\author{Jeff Murugan}
    \email{jeff.murugan@uct.ac.za}
    \affiliation{The Laboratory for Quantum Gravity \& Strings, Department of Mathematics and Applied Mathematics, University of Cape Town, South Africa }
    \affiliation{The National Institute for Theoretical and Computational Sciences, Private Bag X1, Matieland, South Africa}

\author{Jan Olle}
    \email{jan.olle@mpl.mpg.de}
    \affiliation{Max Planck Institute for the Science of Light, Erlangen, Germany}

\author{Dario Rosa}
    \email{dario\_rosa@ictp-saifr.org}
    \affiliation{\ictpadd}
    \affiliation{\pcsadd}
    \affiliation{\ustadd}

\author{Ruth Shir}
    \email{ruth.shir@uni.lu}
    \affiliation{Department of Physics and Materials Science, University of Luxembourg, L-1511 Luxembourg}

\date{\today}

\begin{abstract}
    A deep understanding of the mechanisms underlying many-body quantum chaos is one of the big challenges in contemporary theoretical physics.
    We tackle this problem in the context of a set of perturbed quadratic Sachdev-Ye-Kitaev (SYK) Hamiltonians defined on graphs. 
    This allows us to disentangle the geometrical properties of the underlying single-particle problem and the importance of the interaction terms, showing that the former is the dominant feature ensuring the single-particle to many-body chaotic transition. 
    Our results are verified numerically with state-of-the-art numerical techniques, capable of extracting eigenvalues in a desired energy window of very large Hamiltonians. 
    Our approach essentially provides a new way of viewing many-body chaos from a single-particle perspective.   
\end{abstract}

\maketitle

\section{Introduction}
\label{sec:intro}

After a mild hiatus in the 1990's, \emph{quantum chaos} has reemerged as an extremely active area of research with broad-ranging impact on areas as diverse as black hole physics and strange metals.
The field itself started as early as the very beginning of quantum mechanics, and reached a crescendo with the ground-breaking Bohigas-Giannoni-Schmit conjecture (BGS) of 1984~\cite{bohigas1984chaotic, casati1980connection}, which established that quantum chaotic systems display correlations among their energy levels which are reproduced by the energy levels sampled from random Hamiltonians, the latter being described by random matrix theory (RMT)~\cite{mehta2004random}.
In turn, this equivalence has led to the notion of \emph{universality}, e.g. that many dynamical features (and in particular the long-time features that remain after the transient dynamics have reached completion) of quantum chaotic systems are independent of the fine details of the specific system under investigation. 
The BGS conjecture has been tested in a multitude of settings, most notably billiard systems, and as such is, by now, regarded as a definition of quantum chaos, for \emph{single-particle} (or, at least, few-body) systems.
The reader is referred to Refs.~\onlinecite{10.5555/1214825} and \onlinecite{guhr1998random-matrix} for an extensive discussion of quantum chaos in single-particle systems.

Quantum \emph{many-body} systems, \textit{i.e.} systems built out of many degrees of freedom, are notably more subtle.
For instance, most of the many-body systems of relevance in current experimental setups (such as systems of spin \(1/2\) particles or qubits) \emph{do not} have a classical limit, and therefore talking about quantum chaos is problematic since there is no classical chaotic system to compare with, nor a quantum to classical transition to be inspected.
Here, it is often more useful to talk about \emph{thermalizing} many-body systems, and in the context of the Eigenstate Thermalization Hypothesis (ETH)~\cite{Deutsch1991Quantum, Srednicki1994Chaos} establish a connection between thermalization properties and RMT correlations between the many-body energy levels~\cite{dalessio2016from}.

Another source of complication is that a many-body system, built out by putting together many \emph{non-interacting} single-particle systems, will always be \emph{non-chaotic} from the many-body point of view, irrespective of how chaotic its single-particle components are.
The reason for such a property relies on the many-body energy levels being built as extensive superpositions of the single-particle energy levels, thus removing any correlations that might be present among the single-particle levels.
With this in mind, it becomes interesting to investigate the relationship between single-particle and many-body chaos, since the two phenomena are related but not strictly equivalent.
In turn, this investigation amounts to establishing the role of the interaction in creating many-body RMT correlations.
This set of questions has been addressed in recent years, most notably in Refs.~\onlinecite{winer2020exponential, liao2020many-body, liao2022universal}, where the authors studied the behavior of a typical RMT observable, the Spectral Form Factor (see Sec.~\ref{sec:SFF} for its definition) in a system of non-interacting chaotic particles.
Their results show that the single particle correlations still impact the many-body statistics, by creating an ``exponential ramp'', rather than linear, in the many-body Spectral Form Factor.

In this paper, we take another step in the direction of uncovering the connection between single-particle and many-body chaos.
We consider a set of single-particle systems, defined on graphs of varying connectivity and geometry.
These single-particle systems are then deformed to make them ``minimally many-body'' by adding a \emph{single} quartic term in their Hamiltonians, thus realizing a sort of interacting impurity in an otherwise non-interacting system.
Within this setup, we study how and under what conditions this single impurity is sufficient to induce the many-body energy levels to exhibit RMT-like correlations.

Surprisingly, our numerical results, based on large-scale sparse diagonalization techniques, show that this single impurity can indeed be sufficient to induce RMT-like correlations in the many-body energy levels -- provided that the single particle graph is sufficiently \emph{non-local}.
When instead the single particle graph has a strong local geometry, the many-body energy levels turn out to be non-RMT correlated. 
All in all, our results point toward the importance of single-particle physics in developing many-body quantum chaos.
The role played by the interaction term appears to be subdominant, serving primarily as a single spot where single-particle energy levels can hybridize and mix to develop many-body RMT correlations.
We will also provide a preliminary analysis of the chaotic properties of the underlying single-particle system.
From here it will be clear that the system features a chaotic/non-chaotic single-particle crossover, as a function of the geometry of the graph.
This observation leads us to conclude that even a \emph{single impurity} is enough to develop many-body quantum chaos provided that the underlying single-particle problem is chaotic in the single-particle sense.

The paper is organized as follows.
In Sec.~\ref{sec:model} we introduce the systems under investigation, their main features, and the graphs used to define the single-particle setup.
In Sec.~\ref{sec:r-ratios} we show, using \emph{short range} RMT observables, that the systems under investigation display a chaotic/integrable crossover as a function of the geometry of the underlying graph.
We also present an estimate of the fate of this crossover in the thermodynamic limit, based on a finite-size scaling analysis.
In Sec.~\ref{sec:SFF} we present an analysis of the same transition but this time based on the Spectral Form Factor, which probes \emph{long-range} correlations.
We compare and contrast the results of this analysis with the analysis performed in Sec.~\ref{sec:r-ratios}.
Finally, in Sec.~\ref{sec:discussion}, we present a heuristic explanation of our findings, together with an outlook on possible further developments.

\section{SYK with a single impurity: from single particle to many-body}
\label{sec:model}

Sachdev-Ye-Kitaev (SYK) models of disordered Majorana fermions have emerged in the last few years as a useful set of 
toy models to investigate many-body quantum chaos related questions~\cite{maldacena2016remarks, chowdhury2022Sachdev-Ye-Kitaev, garcia-garcia2016spectral, cotler2017black, kitaev2018the}.

In this context, we inspect how \emph{many-body} quantum chaotic properties can emerge from a minimal modification of a \emph{single particle} problem.
To this end, we consider the following Hamiltonian
\begin{gather}
    \hH = - \mathrm{i} \sum_{G(E,V)} J_{ij} \hat{\gamma}^i \hat{\gamma}^j + \hat{\gamma}^1 \hat{\gamma}^2 \hat{\gamma}^3 \hat{\gamma}^4~, 
    \label{eq:SYK2_4_Hamiltonian}
\end{gather}
where a single particle problem defined on a graph and encoded in a disordered adjacency matrix \(J_{ij}\), is perturbed by \emph{a single} quartic term, that we will refer to as an \emph{impurity}.
The operators \(\hat{\gamma}^i\), with \(i = 1, \dots, N\), are Majorana fermions, satisfying \(\{ \hat{\gamma}^i, \hat{\gamma}^j\} = \delta^{ij}\).
The graph \(G(E,V)\) consists of a collection of \(N\) vertices \(V\) and edges \(E \subseteq V\otimes V\) and it encodes the connectivity of the quadratic terms in Eq.~\eqref{eq:SYK2_4_Hamiltonian} via the following prescription.
We assign a Majorana operator \(\hat{\gamma}^i\) to each vertex \(i \in V\), and a coupling \(J_{ij}\) to each edge \((i,j)\).
The couplings \(J_{ij}\) are random variables sampled from a Gaussian distribution with vanishing mean and variance \(\langle J^2_{ij}\rangle = (N-1)/2 n_E\), where \(n_E\) is the number of edges in \(G(E,V)\).
To make \(\hat{H}\) Hermitian we impose \(J_{ij} = - J_{ji}\).

\begin{figure*}[t!]
\begin{center}
    \includegraphics[width=0.9\hsize]{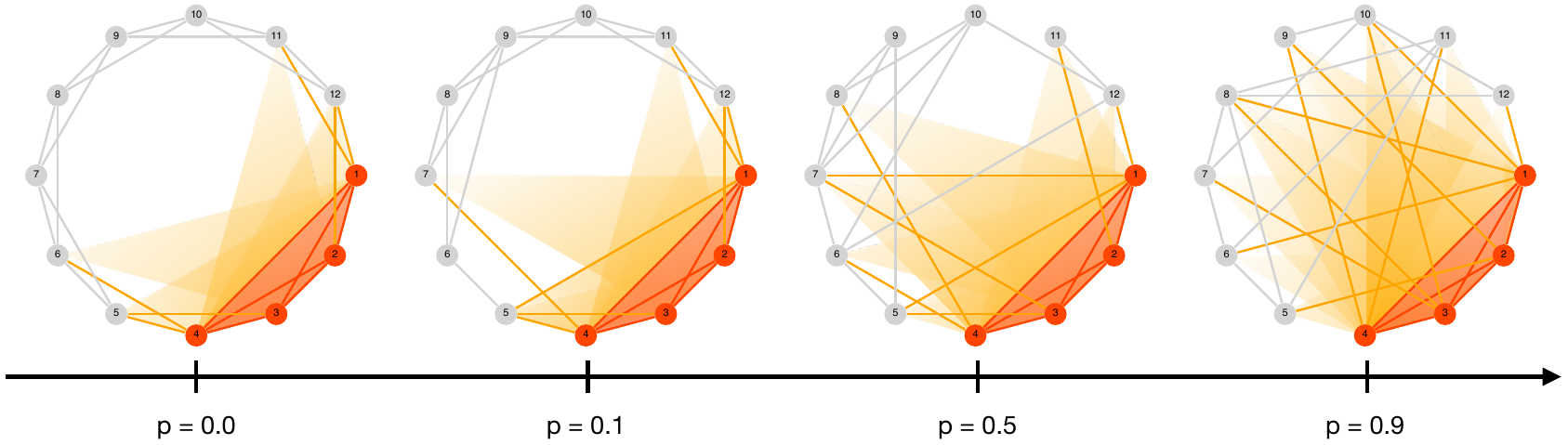}\\
    \caption{
        A pictorial description of the Hamiltonian \eqref{eq:SYK2_4_Hamiltonian} with \(N=12\) Majorana fermions.
        On top of the 2-body interactions described by Watts-Strogatz graphs with \(k=2\) and rewiring probability \(p\)(grey edges), the Majorana fermions 1, 2, 3 and 4 interact via a 4-body interaction (red); the orange edges show the links between the interacting fermions and the rest of the system.
        The effect of this single 4-body interaction is enhanced as the rewiring probability is increased (orange shading).
    }
    \label{fig:graphs_low_large_p_comparison}
\end{center}
\end{figure*}

Among all possible graphs \(G\), we choose \emph{small-world} graphs~\cite{watts1998collective,barabasi}, since it was shown in Ref.~\onlinecite{kim2022operator} that certain dynamical features, like the operator delocalization, for the quadratic part of the Hamiltonian in Eq.~\eqref{eq:SYK2_4_Hamiltonian} is sensitive to the parameters \(k\) and \(p\) defining this family of graphs.
More precisely, these parameters are an integer \(k\) and a probability \(p \in [0,1]\). 
Small-world graph samples are generated by the Watts-Strogatz algorithm~\cite{watts1998collective} which works as follows: 
for a given value of \(k\), start by generating a regular circulant network in which each vertex is connected to its \(2k\) nearest neighbors.
Edges are then randomly rewired with probability \(p\), keeping the graph connected and avoiding self-loops and edge duplications \footnote{It should be emphasized that in Ref.~\onlinecite{watts1998collective}, the constraint on the graph remaining connected after the rewiring procedure \textit{is not} imposed. For this reason, the Newman-Watts variant of the algorithm \cite{newman19993renormalization} is sometimes preferred in the literature on small-world graphs. In this study, we have generated the graphs using the Julia package \texttt{Graphs.jl} \cite{Graphs}, which provides a routine to generate graphs according to the Watts-Strogatz algorithm but with the extra constraint of keeping the graph connected.}.
Importantly, the number of edges in the graph depends on \(k\) only, \(n_E = k N\), while its locality properties are controlled by \(p\), interpolating between a regular lattice (for  \(p=0\)) and a random Erd\"os-Renyi graph~\cite{Erdos1959on, gilbert195random} (for \( p \approx 1 \)).
In the following, we fix \(k = 2\), to keep the model as sparse as possible while maintaining local loops.
We then parametrically vary \(p\) from \(0\) to \(1\).

The role of the two types of terms appearing in Eq.~\eqref{eq:SYK2_4_Hamiltonian} is quite different.
The quadratic term is single-particle in nature.
Very much like the standard \(\mathrm{SYK}_2\) Hamiltonian, it is defined starting from the Hermitian disordered adjacency matrix \(-\mathrm{i} J_{ij}\) which, per se, can be viewed as defining a disordered single-particle quantum problem living on the graph \(G(E,V)\).
This problem is Dyson-like, \textit{i.e.} the disorder is on the hopping terms~\cite{theodorou1976extended,eggarter1978singular,fleishman1977fluctuations,krishna2021beyond}, the on-site energies are all vanishing since \(J_{ii} = 0\), and an extra layer of disorder is given by the random geometry of the graph.
The latter is the only tunable disorder in this setup.
The quadratic term is completed through the products \(\hat{\gamma}^i \hat{\gamma}^j\).
Notice that, given the Majorana nature of the operators \(\hat{\gamma}^i\), the quadratic Hamiltonian \(\hat{h}_{\mathrm{single}} = -\mathrm{i} J_{ij} \hat{\gamma}^i \hat{\gamma}^j\) \emph{does not} preserve the number of particles, while preserving the parity of the number of particles.
By a slight abuse of terminology, we can therefore see \(\hat{h}_{\mathrm{single}}\) as a \emph{trivial} embedding of a single-particle Dyson Hamiltonian into a many-body setting.
However, this embedding must be considered trivial since the many-body energy levels are just obtained as linear combinations of the single-particle energy levels defined by the Dyson problem and no interaction is present.
Therefore, they are always non-chaotic from the many-body point of view, \textit{i.e.} they cannot show level repulsion and/or spectral rigidity.

A genuine many-body picture thus arises via the minimal inclusion of the single term \(\hat{\gamma}^1 \hat{\gamma}^2 \hat{\gamma}^3 \hat{\gamma}^4\) only.
This term, being quartic, induces a mixing between the single-particle energy levels.
However, this mixing is minimal, since it takes place on \(4\) vertices of \(G(E, V)\) only, no matter how large the underlying single-particle problem is, or in other words how large \(N\) is.
See Figure~\ref{fig:graphs_low_large_p_comparison} for a pictorial description of the effect of this 4-body interaction on the system. 
Our goal will be to investigate if, and under which conditions, this minimal interacting term is enough to induce \emph{many-body} quantum chaos among the energy levels.

To compare with recent literature, we note that 
Refs.~\onlinecite{santos2004integrability, torres-herrera2014local, brenes2018high-temperature, brenes2020eigenstate, bastianello2019lack} consider a \emph{many-body integrable} Hamiltonian perturbed by a single impurity term, like the quartic term of Eq.~\eqref{eq:SYK2_4_Hamiltonian}.
On the other hand, Ref.~\onlinecite{bulchandani2022onset} considers a \emph{single particle} Hamiltonian, similar to the quadratic terms of Eq.~\eqref{eq:SYK2_4_Hamiltonian}, perturbed by an \emph{extensive} many-body Hamiltonian. Finally, a dual situation, in which an extensive many-body Hamiltonian is (periodically in time) perturbed by a single-particle Hamiltonian, has been considered in Refs.~\onlinecite{roy2020random, roy2022spectral}.
Our setup differs conceptually from all the aforementioned situations: the extensive terms, being just quadratic, define a \emph{single particle} problem.
This single particle disordered problem is then perturbed by a \emph{single} many-body impurity.
Additionally, it should be stressed that this impurity \emph{is not} perturbative, in the sense that its magnitude is of the same order as the \emph{many} extensive single particle terms.
This setup will allow us to understand how single-particle chaos can be embedded into -- and induces -- many-body chaos.

\section{The r-ratios}
\label{sec:r-ratios}

As a first measure to determine the chaoticity of the many-body energy levels, we compute the well-known \emph{r-ratios}~\cite{atas2013distribution}. 
These quantities can be computed starting from the many-body energy levels, ordered in ascending sense, \(E_1 < E_2 < E_3, \dots\) \footnote{Let us recall that when studying quantum chaotic quantities it is always important to separate the spectrum in symmetry sectors and analyze the data in a \emph{single symmetry sector at a time}.
As already mentioned, the SYK Hamiltonian of Eq.~\eqref{eq:SYK2_4_Hamiltonian} preserves the \emph{parity} of the number of particles appearing in the given Fock representation, \textit{i.e.} states having an odd number of particles are separated from states having an even one~\cite{garcia-garcia2016spectral}.
In this paper, we have systematically analyzed the even symmetry sector only.
The results are of course independent of this choice.}.
From them, one computes the corresponding level spacings \(s_1 = E_2 - E_1\), \(s_2 = E_3 - E_2\), \(\dots\) and finally one obtains the r-ratios
\begin{gather}
    \label{eq:r-ratios_def}
    r_i \equiv \frac{\mathrm{min}(s_{i + 1}, \, s_i)}{\mathrm{max}(s_{i + 1}, \, s_i)} \, .
\end{gather}
In Ref.~\onlinecite{atas2013distribution} it was shown that the averaged \(r\)-ratios take different specific values for spectra that are RMT correlated and for uncorrelated spectra.
Namely, denoting with \(\rr\) the averaged (over disorder realizations) value of the \(r\)-ratios, we have \(\rr \sim 0.60266\) for the case of chaotic spectra following Gaussian Unitary Ensemble correlations (like in the case of the Hamiltonian of Eq.~\eqref{eq:SYK2_4_Hamiltonian}, where the presence of the quadratic term breaks time-reversal symmetry), and \(\rr \sim 0.38629\) for the case of non-chaotic Hamiltonians showing energy levels with Poissonian correlations.

The \(r\)-ratios enjoy properties that make them a very useful tool to probe short-range (in energy scale of the order of the mean-level spacing) chaotic correlations in numerical studies.
In detail, they do not require any unfolding of the energy levels with respect to the mean level density.
Moreover, they are very \emph{intensive} quantities, \textit{i.e.} for a given Hamiltonian \(\hH\) they can be computed from a very minimal number of adjacent eigenvalues (at least \(3\)), close to a certain energy target \(\Bar{E}\).
The results can then be averaged over many realizations of the disordered couplings \(J_{ij}\) and graph realizations.

\begin{figure}[t!]
\begin{center}
    \includegraphics[width=1.0\hsize]{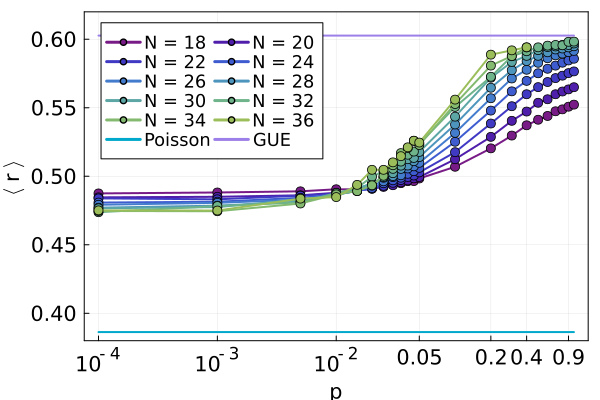}\\
    \caption{
        The transition of the r-ratios values, \(\rr\), as a function of the rewiring probability \(p\) of the underlying graph, computed for the Hamiltonian \(\hH\) of Eq.~\eqref{eq:SYK2_4_Hamiltonian}.
    }
    \label{fig:r_ratios_total}
\end{center}
\end{figure}

These properties make the \(r\)-ratios suitable quantities to be studied via \emph{sparse diagonalization algorithms} and are often used in large diagonalization studies.
In this paper, we have used Chebyshev-filtered Lanczos algorithms (introduced in Ref.~\onlinecite{nyman2001iterative, fang2012filtered, PIEPER2016226} and implemented in the Julia language in Ref.~\onlinecite{ChebyshevFiltering}) to numerically obtain around \(300\) eigenvalues, for each ensemble realization of the Hamiltonian of Eq.~\eqref{eq:SYK2_4_Hamiltonian}, close to the target energy \(\Bar{E} = 0\), \textit{i.e.} in the center of the band \footnote{Our algorithm is very similar, but not fully equivalent, to the POLFED algorithm of Ref.~\onlinecite{sierant2020polynomially}}.
Using these techniques we computed the \(r\)-ratios for several values of the rewiring probability \(p\) and for system sizes from \(N = 18\) up to \(N = 36\) Majorana fermions.
The results, averaged over at least \(1000\) ensemble realizations for the largest \(N = 36\), are shown in Fig.~\ref{fig:r_ratios_total}.

\begin{figure*}[t!]
    \centering
    \begin{subfigure}
        \centering
        \includegraphics[height=2.32in]{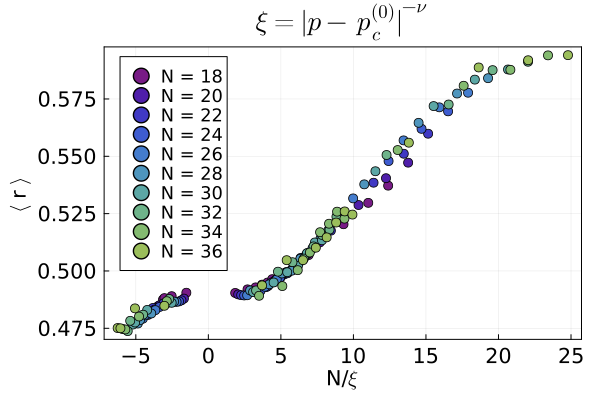}
    \end{subfigure}
    \medskip
    \begin{subfigure}
        \centering
        \includegraphics[height=2.32in]{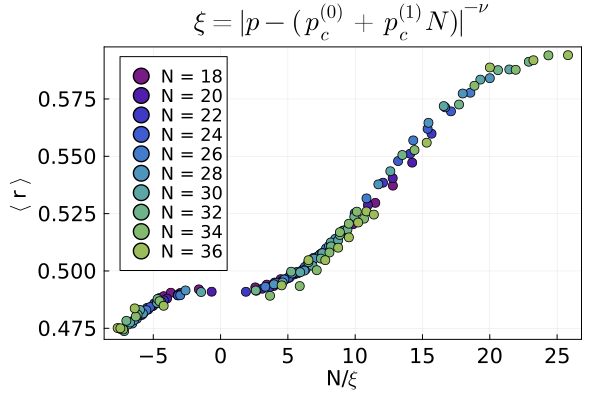}
    \end{subfigure}
    \caption{
        The data collapse computed for the r-ratios \(\langle r  (N, p)\rangle\) plotted against the variable \(N/\xi\) (or \(- N/\xi\) when $p$ is smaller than the critical probability), with the free parameters optimized via the cost-function method.
        Left panel: the minimization is performed via the fixed disorder Ansatz (\(\xi_1\) in Eq.~\eqref{eq:correlation_lenght_ansatz}).
        The critical exponent is \(\nu \approx 0.35647\), the parameter \(p_c^{(0)}\) is \(p_c^{(0)} \approx 0.012340\) and the minimal value of the cost function is \(\mathcal{C} \approx 1.65044\).
        Right panel: the minimization is performed via the drifting disorder Ansatz (\(\xi_2\) in Eq.~\eqref{eq:correlation_lenght_ansatz}).
        The critical exponent is \(\nu \approx 0.35072\), the parameter \(p_c^{(0)}\) is \(p_c^{(0)} \approx 0.02974\) while \(p_c^{(1)}\) reads \(p_c^{(1)} \approx -0.00048\), thus resulting in values very similar for the critical probability to the value found within the fixed disorder ansatz.
        The minimal value of the cost function is \(\mathcal{C} \approx 1.62143\).
    }
    \label{fig:data_collapse}
\end{figure*}

From the plot, we can see that \(\rr\) undergoes a crossover from values well below the GUE value, for very low values of \(p\), to approximately the GUE value, for large enough \(p\).
Moreover, the crossover becomes steeper by increasing the system size \(N\).
Finally, we can see that the crossing point of the curves happens for \(p \approx 0.01\) and no clear drifting of the crossing point is visible.
These two observations seem to suggest that a genuine transition, rather than a smooth crossover, could take place in the thermodynamic limit.
On the other hand, we see that the value of the \(r\)-ratios for very small \(p\) are still far from their expected Poissonian values, with \(\rr \approx 0.478\), though a descending trend with increasing system size is visible.
This indicates that the finite size effects are still strong and that much larger sizes would be necessary to address the behavior of the crossover in the thermodynamic limit.

The transition/crossover can be further investigated by performing a finite-size scaling analysis of the \(r\)-ratios, assuming a single-parameter scaling Ansatz as in the usual 3D Anderson model~\cite{abrahams1979scaling, slevin1999corrections}.
To this end, we have performed a data collapse based on the following Ansatz for the scaling behavior of the \(r\)-ratios
\begin{gather}
    \label{eq;r-ratios_scaling_ansatz}
    \langle r(N, p) \rangle  =  \Theta \left( \mathrm{sign} \left(p - p_c \right) \, \left(\frac{N}{\xi (N, p)} \right)^{1/\nu} \right) \, ,
\end{gather}
where the function \(\Theta\) is unknown, the critical probability is denoted by \(p_c\), the critical exponent is denoted by \(\nu\) (and must be treated as a fitting parameter), and, finally, for the correlation length \(\xi (N, p)\) we have considered the following two power-law Ans\"{a}tze (referred to as \emph{fixed} and \emph{drifting} disorder Ansatz, respectively)
\begin{gather}
    \label{eq:correlation_lenght_ansatz}
    \xi_1  = \frac{1}{\abs{p - p_c^{(0)}}^\nu} , \qquad \xi_2  = \frac{1}{\abs{p - ( p_c^{(0)} + p_c^{(1)} N)}^\nu} \, ,
\end{gather}
with the critical probability parameters \(p_c^{(0)}\) and \(p_c^{(1)}\) that are treated as fitting parameters as well.
The fitting parameters are then determined by the \emph{cost-function} method of Ref.~\onlinecite{suntajs2020ergodicity}.
This method introduces a suitable cost function, \(\mathcal{C}(\rr, \nu, p_c^{(0)}, p_c^{(1)})\), which measures the non-monotonicity of the \(r\)-ratios once ordered according to the value of the variable \(\chi \equiv \mathrm{sign}(p - p_c) \left(N / \xi \right)\).
The values of the fitting parameters are then determined by requiring that they minimize the value of \(\mathcal{C}(\langle r \rangle, \nu, p_c^{(0)}, p_c^{(1)})\), \textit{i.e.} by requiring the best possible monotonicity of \(\rr\) as a function of \(\chi\). 

The results for the data collapse of \(\rr\), for both the fixed and the drifting disorder Ansatz, are reported in Fig.~\ref{fig:data_collapse}.
From the plots, and the corresponding values of the minimized cost function (see the caption of Fig.~\ref{fig:data_collapse}) we see that the two Ans\"{a}tze show quite comparable performances, despite the drifting Ansatz having one extra fitting parameter.
Related to this, the critical rewiring probability, \(p_c\), turns out to be quite insensitive to the system size and it takes the value \(p_c \approx 0.01\), in agreement with the rough position of the crossing point visible in Fig.~\ref{fig:r_ratios_total}.
The weak dependence of the finite-size scaling analysis from the particular form of the scaling Ansatz in Eq.~\eqref{eq:correlation_lenght_ansatz}, is an element that points towards the reliability of the numerical data obtained at finite size and the corresponding finite-size analysis.
On the other hand, we notice that the obtained value of the critical exponent is remarkably low, \(\nu \approx 0.35\).
Such a very low value for the critical exponent is not expected for a many-body system (recall that the Harris bound for the famous disordered Heisenberg model in \(d = 1\) predicts \(\nu \geq 2\)) and it suggests that finite-size effects are strong \footnote{We thank Antonello Scardicchio for discussions on this point.}.
Finally, we note that the finite \(N\) drifting coefficient, \(p_c^{(1)}\), although small is \emph{negative}.
If this trend is not a finite-size effect but a robust phenomenon, it would imply that, in the thermodynamic limit, the critical probability vanishes, \(p_c \to 0\), and that the many-body Hamiltonian \(\hH\) is always chaotic.
Although such a scenario is not likely, and more plausibly the negative drifting of the critical probability is a finite-size effect that disappears in the large-\(N\) limit, our current data does not allow to exclude such a possibility.

\section{The Spectral Form Factor}
\label{sec:SFF}

\begin{figure*}[t!]
    \centering
    \begin{subfigure}
        \centering
        \includegraphics[height=2.32in]{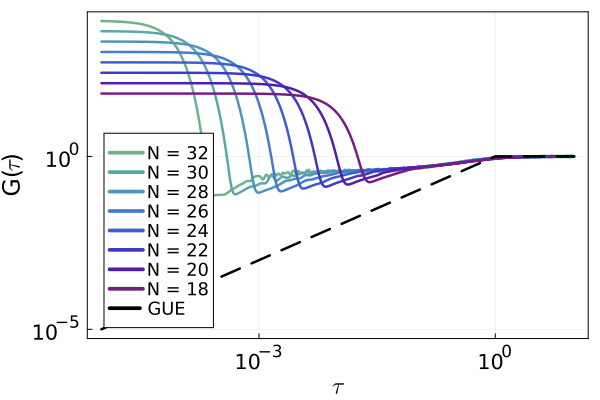}
    \end{subfigure}
    \medskip
    \begin{subfigure}
        \centering
        \includegraphics[height=2.32in]{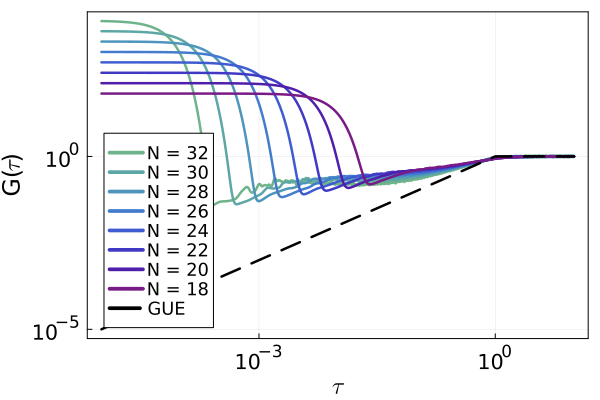}
    \end{subfigure}
    \centering
    \begin{subfigure}
        \centering
        \includegraphics[height=2.32in]{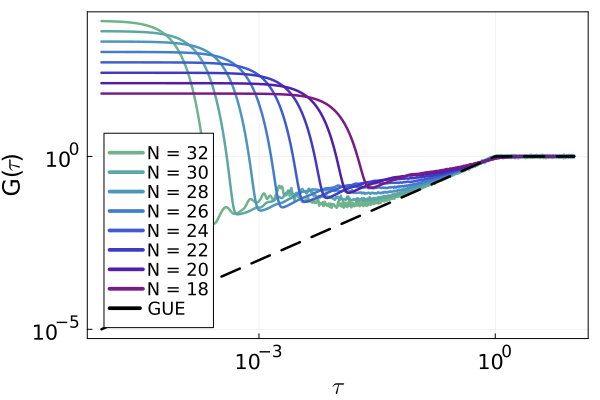}
    \end{subfigure}
    \medskip
    \begin{subfigure}
        \centering
        \includegraphics[height=2.32in]{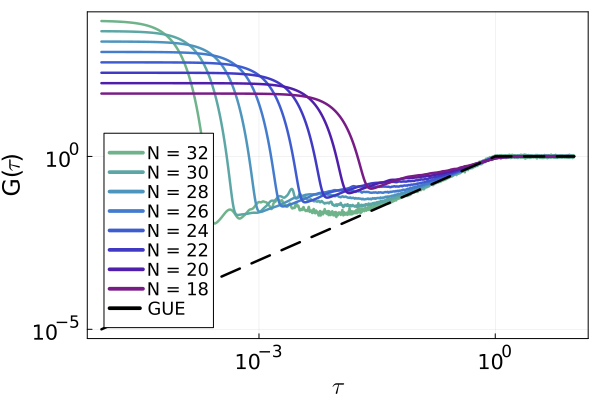}
    \end{subfigure}
    \caption{
        The Spectral Form Factor, computed for system sizes ranging from \(N = 18\) up to \(N = 32\), for the values \(p = 0.0\) (upper left), \(p = 0.1\) (upper right), \(p = 0.5\) (lower left), and \(p = 1.0\) (lower right).
        The ensemble averages are taken over several thousand, at least \(5000\) for \(N = 32\), of disorder and graph realizations.
    }
    \label{fig:SFF}
\end{figure*}

By construction, the \(r\)-ratios probe level correlations at energy scales of the order of the mean level spacing.
On the contrary, they are completely insensitive to the level correlations enjoyed by energy levels which are far apart from each other.
To investigate these aspects, we now focus on the Spectral Form Factor (SFF), for several values of \(p\).
The SFF can be defined as follows~\cite{suntajs2020quantum, nandy2022delayed}
\begin{gather}
    \label{eq:SFF_def}
    G(\tau) \equiv \frac{1}{Z} \left\langle \abs{\sum_{n = 1}^D \rho(\epsilon_n) \, \mathrm{exp} \left( -i 2 \pi \epsilon_n \tau \right)}^2 \right\rangle \, ,
\end{gather}
where the symbol \(\left\langle \cdot \right\rangle\) indicates the average over the Gaussian couplings and the random graphs.
Moreover, we denoted the dimension of the Hilbert space, in the given symmetry sector, as \(D\) (in the case at hand, we have \(D = 2^{\frac{N}{2} - 1}\)), the factor \(Z\) is chosen to ensure \(G(\tau) \to 1\) for \(\tau \to \infty\), the unfolded energy levels are denoted by \(\epsilon_n\), and the function \(\rho(\epsilon_n)\) is a \emph{Gaussian filtering function},
whose role is to suppress the contribution from the edges of the (unfolded) eigenvalue distribution and magnify the contribution from the energy levels lying on the center of the band~\cite{gharibyan2018onset}.
See Ref.~\onlinecite{nandy2022delayed} for the details on the unfolding procedure, the definition of \(Z\), and the definition of the filtering function \(\rho(\epsilon_n)\).

It is easy to show that the SFF computes the Fourier transform of the density-density correlator.
Therefore, it can be used to investigate long-range energy correlations and the presence or absence of \emph{spectral rigidity}~\cite{guhr1998random-matrix}.
In particular, it is possible to extract from it the so-called \emph{Thouless time}, \(\tau_{\mathrm{Th}}\), \textit{i.e.} the timescale (measured in units of the inverse of the mean level spacing, also referred to as the \emph{Heisenberg time}) after which the time evolution of a chaotic quantum system becomes described by Random Matrix Theory.
It should be stressed that the computation of the SFF requires, by definition, the knowledge of the \emph{entire} many-body spectrum.
Therefore, it cannot be computed using sparse diagonalization techniques and requires \emph{full} diagonalization algorithms.

We have computed the SFF for the model under consideration for the values of \(p = 0.0, 0.1, 0.5, 1.0\), and for system sizes ranging from \(N = 18\) up to \(N = 32\).
The results are plotted against the RMT prediction (for the GUE ensemble) in Fig.~\ref{fig:SFF}.

From the plots, we can see that for both \(p=0.0\) and \(0.1\) the SFF \emph{does not} show any RMT-like behavior.
In particular, it does not display the linear ramp dictated by spectral rigidity.
Based on the results of Sec.~\ref{sec:r-ratios}, this result has to be expected for \(p = 0.0\).
On the contrary, it is somehow surprising for \(p = 0.1\), since \(p = 0.1\) is way larger than the critical probability, \(p_c \approx 0.012\), as measured by the \(r\)-ratios.
The conclusion that we can take from this observation is that the system approaches quantum many-body chaotic properties very slowly with increasing \(p\) and that \(p_c \approx 0.012\) is the probability necessary to induce short-range RMT-correlations but much larger probabilities of rewiring are necessary to make these RMT-correlations robust at larger energy scales.

This picture is confirmed by looking at \(p = 0.5\) and \(p = 1.0\).
In this case, the SFF shows the expected ramp-like behavior of GUE.
In particular, we see that \(\tau_{\mathrm{Th}}\) (identified as the time at which the SFF agrees with the GUE curve) becomes smaller and smaller (in units of the Heisenberg time) while increasing the system size, thus showing the ergodicity of the model for these values of \(p\).
We notice that contrary to the standard SYK model (including its sparse generalizations) the SFF shows an intermediate regime, between the end of the initial fall and the linear ramp defined by the Thouless time.
During this time window, the SFF shows a kind of mixed behavior, reminiscent of the analogous behavior displayed by the SFF of the mass-deformed SYK model~\cite{PhysRevLett.120.241603, nosaka2018the, nandy2022delayed} for intermediate values of the mass-deformation strength.
It is interesting to stress that, for the mass-deformed SYK model, this intermediate regime does not exclude the presence (at a sufficiently low magnitude of the mass deformation parameter) of the fully ergodic \(\mathrm{SYK}_4\) SFF, as shown analytically in Ref.~\onlinecite{monteiro2021minimal}.
On the other hand, from the numerical results presented here, we could not find a sufficiently large value of \(p\) for which the SFF becomes fully ergodic.

\section{Discussion and Outlook}
\label{sec:discussion}

To summarize, the results coming from the \(r\)-ratios and the SFF analysis present a clear indication that the many-body model described by Eq.~\eqref{eq:SYK2_4_Hamiltonian} enjoys an integrable/chaotic crossover, as a function of the rewiring probability \(p\) and consequently, the geometry of the graph defining the quadratic part of the Hamiltonian.

Heuristically, this can be explained as follows:
To begin, the quadratic part of the Hamiltonian can be always diagonalized, via a \emph{canonical} Bogoliubov transformation \( \hat{\mathcal{U}} \), to the form
\begin{gather}
    \label{eq:rotated_quadratic_hamiltonian}
    \hat{\Tilde H}_2 = - \mathrm{i} \sum_{i \in \mathrm{odd}} \epsilon_i \hat{\chi}^i \hat{\chi}^{i + 1} \, ,
\end{gather}
where \(\epsilon_i\) are the single particle energy levels.
Since \(\hat{\mathcal{U}}\) is canonical, the new Majorana fermions \(\hat{\chi}^i\) satisfy the standard anticommutation relations \( \left\{\hat{\chi}^i, \, \hat{\chi}^j \right\} = \delta^{ij}\), i.e. they are \emph{as fundamental as} the operators \(\hat{\gamma}^i\).
When expressed in terms of the new fermions \(\hat{\chi}^i\), the dependence of the model on the geometry of the graph is obscured since it is encoded in the single particle energy level \(\epsilon_i\) and the definition of the new Majorana fermions \(\hat{\chi}^i\).

\begin{figure}[t!]
\begin{center}
    \includegraphics[width=1\hsize]{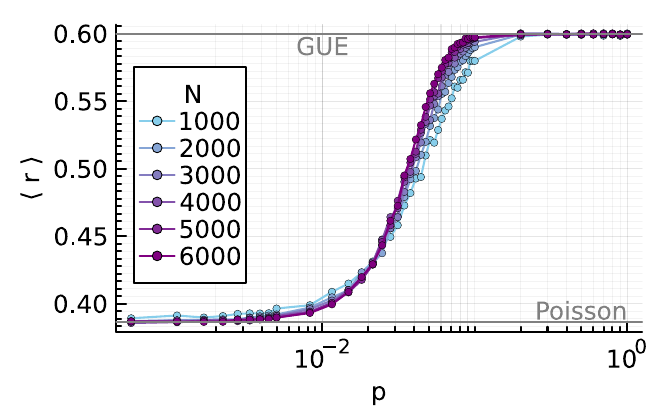}\\
    \caption{
        Transition from Poissonian value of \(\rr\) to GUE value for the coupling matrix \( \mathrm{i}J_{ij}\) as a function of the rewiring probability of the underlying graph.
        For each \(N\) we generated a connected graph with rewiring probability \(p\).
        For that graph we then performed an ensemble average over \(30\) or \(50\) random realizations of the weights \(J_{ij}\).
        We repeated this process at least 10 times for each \(N\) and performed another average.
    }
    \label{fig:r_single_particle}
\end{center}
\end{figure}

To get further insight, it is convenient to consider the \emph{full} Hamiltonian, \(\hat{H}\), in Eq.~\eqref{eq:SYK2_4_Hamiltonian}. 
In terms of the fermions \(\hat{\chi}^i\) it reads
\begin{gather}
    \label{eq:SYK2_4_Hamiltonian_rotated}
    \hH = - \mathrm{i} \sum_{i \in \mathrm{odd}} \epsilon_i \hat{\chi}^i \hat{\chi}^{i + 1} + \hat{\Tilde{H}}_4(\hat{\chi}^i) \, ,
\end{gather}
where the expression \(\hat{\Tilde{H}}_4(\hat{\chi}^i)\) denotes the quartic Hamiltonian obtained by acting with \(\hat{\mathcal{U}}\) on the interacting term \(\hat{\gamma}^1 \hat{\gamma}^2 \hat{\gamma}^3 \hat{\gamma}^4\).

We are then led to conclude that, once rewritten as in Eq.~\eqref{eq:SYK2_4_Hamiltonian_rotated}, the model is many-body chaotic if \(\hat{\Tilde{H}}_4(\hat{\chi}^i)\) displays an \emph{extensive} number of terms.
This conclusion agrees with studies on sparse versions of SYK, which have shown that the details of the sparsification procedure are largely irrelevant as long as an extensive number of terms are preserved~\cite{xu2020a, PhysRevD.103.106002, caceres2021sparse, caceres2022spectral, orman2024quantum}.
Notice, however, that the intermediate-time behavior of the SFF, visible for \(p = 0.5, 1.0\) \emph{is not} visible in the sparse quartic SYK model~\cite{caceres2022spectral, orman2024quantum}, and its origin must be traced back to the presence of the random mass term in Eq.~\eqref{eq:SYK2_4_Hamiltonian_rotated}.

This line of reasoning begs the question: under 
which conditions does the Bogoliubov transformation \(\hat{\mathcal{U}}\) generate an extensive number of quartic terms when acting on \(\hat{\gamma}^1 \hat{\gamma}^2 \hat{\gamma}^3 \hat{\gamma}^4\)?
To answer this, we observe that the columns of \( \hat{\mathcal{U}} \) are given by the eigenvectors, \(\ket{\psi_i}\), of the \emph{single particle} problem defined by the coupling matrix \( \mathrm{i}J_{ij}\).
As already discussed, these can be viewed as a Dyson-like problem defined over a Watts-Strogatz graph. 
Hence, we argue that \(\hH\) is many-body chaotic when the eigenvectors \(\ket{\psi_i}\) are \emph{extended} over the Watts-Strogatz graph, 
since in this case \(\hat{\mathcal{U}}\) will have an extensive number of columns with non-vanishing components over all the vertices of the Watts-Strogatz graph.
This in turn creates an extensive number of terms when acting on \(\hat{\gamma}^1 \hat{\gamma}^2 \hat{\gamma}^3 \hat{\gamma}^4\).
Indeed, the coupling matrix \(\mathrm{i}J_{ij}\) shows an integrable/chaotic transition when \(p\) increases from \(0\) to \(1\), see Fig.~\ref{fig:r_single_particle}, which will be studied elsewhere and looks qualitatively similar to an analogous transition observed in Ref.~\onlinecite{cugliandolo2024multifractal} for Dyson-like models defined on Erd\"os-R\'enyi graphs.

The single-particle perspective could also explain the origin of the large deviations from the expected Poisson value of the \(r\)-ratios, observed for vanishing or small rewiring.
Such nearest-neighbor Dyson-like Hamiltonians on lattices have diverging density of states and localization length at \(E_{\mathrm{sp}}=0\) as well as anomalously localized \(E_{\mathrm{sp}}=0\) states~\cite{Soukoulis_Economou1981, Inui_Trugman_Abrahams1994,detomasi2016generalized}.
We conjecture that these large deviations are due to the presence of these anomalously localized states.
Therefore they are finite-size effects, which are nevertheless very strong and tend to disappear for huge sizes only, which cannot be studied via state-of-the-art exact diagonalization techniques.

Although this heuristic picture is appealing and reasonable, several open issues should be further investigated.
Of course, the most pressing question is to determine the nature of the chaotic/integrable crossover as detected by the \(r\)-ratios.
Is it a genuine transition or a smooth crossover in the thermodynamic limit?
Our results are, in certain respects, contradictory on this point; on the one hand, the data collapses for the drifting v.s. the fixed critical disorder ansatz show a remarkable level of agreement, thus suggesting that finite size effects could be not very strong compared to other situations like, for example, the highly debated transition for many-body localization~\cite{luitz2015many-body, pietracaprina2018shift-invert, suntajs2020quantum}.
On the other hand, the small value of the critical exponent and the intermediate statistics of the r-ratios for \(p \to 0\) suggest that the system could still be very far from a scaling behavior. 
In addition, the data collapse has been realized within the usual assumption that a single-parameter scaling hypothesis (like for the Anderson model at low dimensionality) can be applied.
It has been recently argued in Refs.~\onlinecite{vanoni2024renormalization} and \onlinecite{altshuler2024renormalizationg} that the validity of this assumption should not be taken for granted and that a more cautious way to study transitions of the form delocalization/localization is via the study of the full \(\beta\) function.
It would be of immediate interest to perform such an analysis for this model.

Another potential issue worth further investigation is the very tiny \emph{negative} drifting with the system size of the critical disorder that we found by using the drifting disorder Ansatz.
Is this drifting a finite-size effect? a numerical artifact? or a genuine feature of the model?
In this respect, we anticipate that answering this question would bring immediate consequences for the long-standing debate on the fate of many-body localization in the thermodynamic limit. Eq.~\eqref{eq:SYK2_4_Hamiltonian} indeed describes a single particle problem with a single interacting site (For other situations where a single impurity can change dramatically the physical properties of the system under investigation we refer to Refs.~\onlinecite{brighi2022localization}, \onlinecite{gao2024information}, and \onlinecite{molignini2023anomalous}).
Since from Fig.~\ref{fig:r_single_particle} we know that the single particle problem has a localized regime, it is evident that this many-body problem can be viewed as a very minimal version of the standard situation studied in many-body localization.
Indeed, Eq.~\eqref{eq:SYK2_4_Hamiltonian_rotated} represents the starting point for the study of many-body localization as a localization problem in Fock space, rather than in real space, as advocated starting from Ref.~\onlinecite{altshuler1997quasiparticle}.
It would be very interesting to show that, at least for this very minimal setup in which the interaction is happening on a single site, localization is robust against the presence of interactions, and this negative drifting is just a finite-size effect. 

Finally, it will also be interesting to explore further the thermalization properties of this model, and fully characterize the intermediate regime displayed by the SFF.
In this respect, we believe that the analytical methods developed very recently in Ref.~\onlinecite{altland2024tensor} will be very useful.

\section*{Acknowledgements}
\label{sec:acknowledgements}

We thank Boris Altshuler, William Salazar, Antonello Scardicchio, Marco Tarzia, and Davide Venturelli for the interesting discussions. 
AA and DR acknowledge the support from the Institute for Basic Science in Korea (IBS-R024-D1). 
DR thanks FAPESP, for the ICTP-SAIFR grant 2021/14335-0 and for the Young Investigator grant 2023/11832-9.
DR also acknowledges the Simons Foundation for the Targeted Grant to ICTP-SAIFR.
JO thanks PCS-IBS for hospitality through the PCS-IBS Visitors Program.
JO's work is supported by the Munich Quantum Valley, which is supported by the Bavarian state government with funds from the Hightech Agenda Bayern Plus. 
JM acknowledges support from the ``Quantum Technologies for Sustainable Development'' grant from the National Institute for Theoretical and Computational Sciences of South Africa (NITHECS).
AA, MC, and DR acknowledge the MOU agreement between IBS and CNR-SPIN.

\end{document}